\documentclass[twocolumn,prl,aps,amsmath,amssymb,superscriptaddress]{revtex4-2}

\usepackage{graphicx}
\usepackage[colorlinks=true,allcolors=blue]{hyperref}
\usepackage{xcolor}
\usepackage{lipsum}
\usepackage{siunitx}
\usepackage{etoolbox}
\usepackage[toc,page]{appendix}
\usepackage[capitalise]{cleveref}

\crefname{appendix}{App.}{Apps.}
\Crefname{appendix}{Appendix}{Appendices}

\makeatletter
\AtBeginEnvironment{appendix}{%
    \crefalias{section}{appendix}%
}
\makeatother

\begin{document}

\title{Improving Neutrino Point Source Sensitivity with Source-Informed Event Selection}
% \title{From Neutrino Observatories to Neutrino Telescopes:\\ Improving Neutrino Point Source Sensitivity with Source-Informed Event Selection}

\author{Jeffrey Lazar}
\email{jlazar@icecube.wisc.edu}
\affiliation{Centre for Cosmology, Particle Physics and Phenomenology -- CP3, Universit\'{e} catholique de Louvain, Louvain-la-Neuve, Belgium}

\author{Carlos A. Arg\"{u}elles}
\email{carguelles@g.harvard.edu}
\affiliation{Department of Physics \& Laboratory for Particle Physics and Cosmology, Harvard University, Cambridge, MA 02138, USA}

\author{Pavel Zhelnin}
\email{pzhelnin@g.harvard.edu}
\affiliation{Department of Physics \& Laboratory for Particle Physics and Cosmology, Harvard University, Cambridge, MA 02138, USA}
\date{\today}

\begin{abstract}
Neutrino telescopes employ multi-level reconstruction chains, where computationally expensive high-quality reconstructions are applied only to events that survive initial quality cuts based on fast, coarse directional estimates.
Currently, event selection between reconstruction levels is source-agnostic, giving no priority to events from directions of known neutrino source candidates.
We propose a simple modification to inter-level event selection: preferentially retain events whose early-level reconstruction places them within an angular tolerance of pre-specified candidate source directions from established multi-messenger catalogs, while continuing to subsample remaining events at the baseline rate.
Using a realistic two-level detector model with energy-dependent angular resolution, we show that this source-informed selection can improve median point source sensitivity by factors of $\sim 2$--$3$ compared to uniform subsampling, with the improvement depending on the baseline selection efficiency, angular tolerance, and correlation between reconstruction qualities at different levels.
For catalogs of $\mathcal{O}(100)$ sources, the additional computational overhead is modest ($\sim 7$--$14\%$).
This approach offers a path to substantially enhance the discovery potential of current and future neutrino telescopes without requiring new detector capabilities.
\end{abstract}

\maketitle

\textbf{\textit{Introduction---}} In the decade-plus since its completion, the IceCube Neutrino Observatory, the first gigaton-scale neutrino telescope, has opened a window to the neutrino sky.
The experiment has achieved its central science goals by discovering the diffuse astrophysical neutrino flux and identifying astrophysical neutrino point sources, including neutrino emission associated with the blazar TXS~0506+056~\cite{IceCube:2018dnn}, the Seyfert galaxy NGC~1068~\cite{IceCube:2022der}, and the
Galactic plane~\cite{IceCube:2023ame}; see
Ref.~\cite{Arguelles:2024ncf} for a recent review.

Additionally, a new generation of gigaton-scale detectors is coming online and is already contributing to the study of the neutrino sky.
Notably, the Baikal-GVD detector~\cite{GVD:2025lya} has independently measured the all-sky astrophysical neutrino flux, and the KM3NeT detector~\cite{KM3NeT:2025npi} recorded the highest-energy fundamental particle ever observed, which challenges our understanding of the supra-PeV neutrino flux~\cite{KM3NeT:2025ccp,Li:2025tqf,Yuan:2025isv}.
These detectors will be further complemented by other, next-generation, planned detectors, such as IceCube Gen2, P-ONE~\cite{P-ONE:2020ljt}, TRIDENT~\cite{TRIDENT:2022hql}, and HUNT~\cite{Huang:2023mzt}.

Despite the early successes and future promise of these experiments, point source searches remain a challenge.
The known extragalactic neutrino sources can comprise only a small fraction of the observed diffuse neutrino flux~\cite{IceCube:2022der,IceCube:2023ame}.
And since these analyses are background-dominated, the additional exposure brought by the current and planned telescopes will only improve our source sensitivity by a factor of a few by 2040~\cite{Schumacher:2025qca}, assuming optimistically that all proposed experiments are built.
This necessitates revisiting and reevaluating our point-source searches and potentially designing new approaches to accelerate our discovery.

\begin{figure}[b]
    \centering
    \includegraphics[width=\columnwidth]{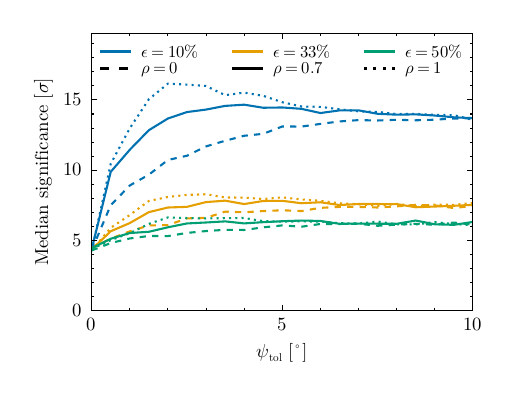}
    \caption{\textbf{\textit{Median expected significance as a function of angular tolerance.}}
    The three colors indicate different selection efficiencies $\epsilon = 10\%$, $33\%$, and $50\%$, while the line styles indicate the inter-level quality correlation: dashed ($\rho=0$), solid ($\rho=0.7$), and dotted ($\rho=1$).
    Our procedure saves more events for more aggressive cuts, resulting in larger improvements in those cases.
    Inter-level quality correlations result in a larger derivative and, in the most extreme case considered, a local maximum.
    }
    \label{fig:significance}
\end{figure}

Examining the data-processing chains employed in these experiments to remove unwanted background, typically atmospheric muons, offers such an opportunity.
These processing chains are organized into successive reconstruction levels~\cite{AMANDA:2003vtt} of increasing computational cost and improved precision.
Early levels apply fast but coarse directional reconstructions to the full event stream~\cite{Aartsen:2013bfa}, followed by quality cuts that reduce the sample to a manageable size for more sophisticated---and computationally expensive---algorithms at later levels~\cite{IceCube:2021oqo,IceCube:2023avo}.
Crucially, events that do not survive the intermediate selection never receive the higher-quality reconstruction, regardless of their potential scientific value.

We propose a simple modification: events whose early-level reconstruction falls within an angular tolerance of a candidate source direction are preferentially retained for higher-level processing, while the remaining events continue to be subsampled at the baseline rate.
The intuition is straightforward---if an event's coarse reconstruction places it near a known source, that event is more likely to be an astrophysical signal and would benefit disproportionately from a higher-quality reconstruction.
While this direction-agnostic approach was natural when no sources had been identified, the growing catalog of neutrino source candidates and the near-future prospect of inter-observatory searches motivate reconsidering this default.
Interestingly, relaxing cuts when an external constraint suppresses backgrounds has already been shown to enhance sensitivity in the time domain~\cite{Fahey:2018thesis}; however, the logic has not been extended to the spatial domain.

In this Letter, we use a realistic two-level detector model to demonstrate that this \emph{source-informed event selection} can substantially improve point-source sensitivity---by factors of $\sim 2$--$3$ in median significance---with modest additional computational overhead.

\textbf{\textit{Simulation Setup and Analysis Method---}} We consider a neutrino telescope that provides two independent directional reconstructions per event, labeled $L_1$ and $L_2$, with energy-dependent angular resolutions $\sigma_1(E)$ and $\sigma_2(E)$.
The angular errors $\kappa_{i}$ are drawn from half-normal distributions $\kappa_i \sim \mathrm{HalfNormal}(\sigma_i)$, and the correlation between the two reconstruction qualities---i.e. the quantile of the event angular error---is parameterized by $\rho \in [0,1]$ via a Gaussian copula.
When $\rho = 0$ the errors are independent; when $\rho = 1$, they are identical; intermediate values interpolate between these limits.

\begin{figure}[t]
    \centering
    \includegraphics[width=\columnwidth]{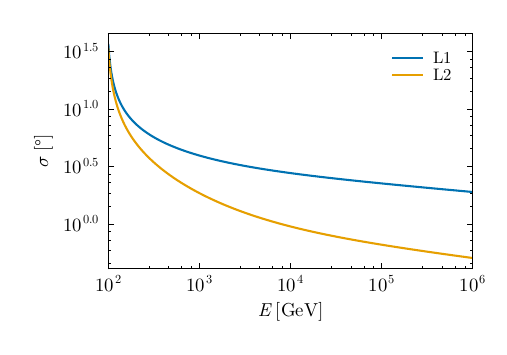}
    \caption{\textbf{\textit{Angular resolution as a function of neutrino energy.}}
    The heuristic function captures important facets of the neutrino telescopes.
    First, a divergent regime at the lower energies, followed by a regime scaling like $E_{\nu}^{0.7}$ where the muon kinematic angle limits the reconstruction, and finally a reconstruction-limited regime that improves less quickly.
    }
    \label{fig:resolution}
\end{figure}

For concreteness, we adopt a detector model with energy-dependent resolution given by:
$$
\sigma_{i}(E) = \frac{\sigma^{(0)}}{(E - E^{(0)})^{p^{(0)}}} + \frac{\sigma_{i}^{(1)}}{(E - E^{(0)})^{p^{(1)}_{i}}},
$$
with $\sigma_0 = 100^{\circ}$, $E_{0}=\SI{95}{\GeV}$, and $p_{0}=0.7$ the same between the two levels, and $(\sigma_{1}^{(1)}, p_{1}^{1})=(5^{\circ}, 0.07)$ and $(\sigma_{2}^{(1)}, p_{2}^{1})=(2^{\circ}, 0.1)$.
While heuristic, this functional form, shown in \cref{fig:resolution} captures important features of angular resolutions in neutrino telescopes, i.e., a quickly improving, kinematic-limited regime at low energies and a more slowly improving, reconstruction-limited regime at higher energies~\cite{AMANDA:2003vtt,IceCube:2021oqo}.

We sample signal events from a power-law spectrum $\propto E^{-\gamma_{\mathrm{sig}}}$ with $\gamma_{\mathrm{sig}} = 3.2$ and are distributed according to the point-spread function centered on a source direction assumed known a priori from catalog or multi-messenger information, while background events are isotropically distributed with spectral index $\gamma_{\mathrm{bg}} = 3.7$.
The signal normalization is tuned so that approximately 87 signal events survive baseline selection at $\psi_{\mathrm{tol}} = 0$, corresponding to a median significance of $\sim 4\,\sigma$, against a background of $\sim 1.4 \times 10^{6}$ selected events.

The central idea is to exploit the angular information from $L_1$ to preferentially retain events consistent with the candidate source direction.
We define a \emph{quasirandom selection} procedure with two parameters: an angular tolerance $\psi_{\mathrm{tol}}$ and a baseline efficiency
$\epsilon$.
If the $L_1$ reconstruction falls within $\psi_{\mathrm{tol}}$ of the source direction, the event is \emph{always} retained.
Otherwise, the event is retained with probability $\epsilon$.

This selection enriches the sample with events whose better reconstruction points toward the source, without discarding events outright.
Events passing the selection are then analyzed using only the $L_2$ reconstruction, which provides an independent directional estimate.

When $\psi_{\mathrm{tol}} = 0$, no angular cut is applied and the selection reduces to uniform random subsampling at rate $\epsilon$.  
As $\psi_{\mathrm{tol}}$ increases, more signal events are captured in the cone, improving sensitivity up to an optimal tolerance beyond which additional background dilutes the gain.

Selected events are binned in $\cos\psi$, where $\psi$ is the angular distance between the $L_2$ reconstruction and the source direction, using $N_{\mathrm{bins}} = 20{,}000$ bins with width $\Delta\cos\psi = 10^{-4}$.
We construct signal and background probability density functions (PDFs) from Monte Carlo simulation.
The signal PDF is generated from $5 \times 10^{5}$ signal events passed through the selection, histogrammed and normalized.
The background PDF is generated by passing $5 \times 10^{7}$ isotropic events through the selection in batches, histogramming them, and normalizing.
This ``direct'' template construction captures both the enhancement of background events within the cone and the corresponding depletion at larger angular distances.

\begin{figure}[t]
    \centering
    \includegraphics[width=\columnwidth]{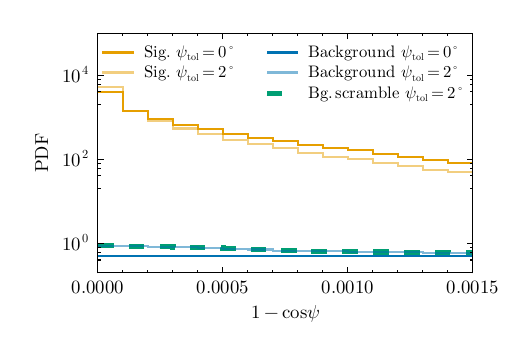}
    \caption{\textbf{\textit{Signal and background PDFs as a function of $1 - \cos\psi$ for $\epsilon = 33\%$ and $\rho = 0.7$}}.
    Solid lines are computed as described in the text.
    The darker lines show the case with no angular selection, i.e. $\psi_{\mathrm{tol}} = 0$, where the background is flat.
    lighter colors show $\psi_{\mathrm{tol}} = 2^\circ$, where the background acquires a cone enhancement near the source and the signal PDF narrows due to the correlated quality selection.
    The green dashed line shows the background PDF computed using the source scrambling method described in the text.
    }
    \label{fig:pdfs}
\end{figure}

\cref{fig:pdfs} illustrates the signal and background PDFs for $\psi_{\mathrm{tol}}=0^{\circ}$ and $\psi_{\mathrm{tol}}=2^{\circ}$.
The quality-dependent selection produces a characteristic background shape: a narrow enhancement near $\cos\psi = 1$ (the cone) superimposed on a depleted bulk distribution.  The signal PDF peaks sharply at $\cos\psi = 1$ and falls off according to the point-spread function of the $L_2$ reconstruction.

Analyzers typically exploit Earth's rotation to construct background distributions.
By assigning a uniformly drawn random time within the analysis window to data events, the relationship between local and sky-fixed coordinates is broken.
This dissociates signal events from the true source direction, yielding a background-free sample and sparing the analyzer the need to generate large simulation sets and contend with systematics.
Our proposed procedure explicitly breaks the symmetry imposed by Earth's rotation, and it selects a preferred point in the sky.
However, we can retain these benefits by randomizing the source right ascension while keeping the events fixed, thereby introducing a time offset between the true direction and the new direction.
\cref{fig:pdfs} shows that this source-randomization procedure---green dashed line, 500 signal events and 500 random right ascension values---agrees excellently with the explicitly calculated background PDF, validating the approach.

The test statistic is a binned Poisson log-likelihood ratio,
\begin{equation}
    \mathrm{TS} = -2 \ln \frac{\mathcal{L}(n_s=0)}
                               {\mathcal{L}(\hat{n}_s)},
    \label{eq:ts}
\end{equation}
where $\hat{n}_s \geq 0$ is the best-fit number of signal events, obtained by maximizing the Poisson likelihood over $n_s$.

The expected number of events in bin $i$ under the signal-plus-background
hypothesis is
\begin{equation}
    \mu_i = N_{\mathrm{bg}}\, b_i\, \Delta\cos\psi
          + N_{\mathrm{sig}}\, s_i\, \Delta\cos\psi,
\end{equation}
where $b_i$ and $s_i$ are the background and signal PDFs evaluated at bin $i$, and $N_{\mathrm{bg}}$, $N_{\mathrm{sig}}$ are the total expected background and signal counts after selection.
As shown in \cref{app:fit_validation}, our background trials follow the expected $\chi^{2}$ distribution and the fitting procedure correctly recovers the injected number of events.

\textbf{\textit{Results---}}We compute the expected median significance across a grid of selection parameters: quality correlation $\rho~\in~\{0, 0.7, 1\}$, baseline efficiency $\epsilon~\in~\{10\%, 33\%, 50\%\}$, and angular tolerance $\psi_{\mathrm{tol}} \in [0^\circ, 10^\circ]$ in steps of $0.5^\circ$.
The median significance $\sigma$ is obtained from the median test statistic over 500 signal-plus-background Poisson realizations per
configuration.

\cref{fig:significance} shows the significance as a function of
tolerance for all configurations.
Several features are evident.
First, At $\psi_{\mathrm{tol}} = 0$ (no quality-based selection), the
significance is $\sim 4\,\sigma$ regardless of $\rho$, as expected.
Secondly, significance rises steeply with tolerance, reaching a broad maximum at $\psi_{\mathrm{tol}} \approx 3$--$8^\circ$ depending on the configuration.
Lastly, Beyond the optimal tolerance, the significance plateaus or gently declines as the cone admits increasing amounts of background.

Higher quality correlation $\rho$ leads to greater improvement from the selection.
When $\rho = 1$ (shared quality), an event with $L_1$ near the source almost certainly has $L_2$ near the source as well, making the cone selection highly efficient at capturing signal.
When $\rho = 0$ (independent qualities), the cone still enriches the sample because $L_1$ provides directional information even though $L_2$ is uncorrelated---but the improvement is smaller.

Lower baseline efficiency $\epsilon$ yields greater \emph{relative} improvement because the cone-selected events constitute a larger fraction of the retained sample.
At $\epsilon = 10\%$, the selected sample is dominated by events whose $L_1$ reconstruction falls near the source, while at $\epsilon = 50\%$, a larger fraction of randomly retained background dilutes the signal enhancement.

The optimal tolerance depends on the angular resolution of the selecting
reconstruction $L_1$: it should be large enough to capture most signal
events whose $L_1$ error falls within the point-spread function, but
small enough to avoid excessive background contamination.  Empirically,
the optimal $\psi_{\mathrm{tol}}$ scales roughly with the median
$\sigma_1$ of the $L_1$ reconstruction.

\begin{figure}[t]
    \centering
    \includegraphics[width=\columnwidth]{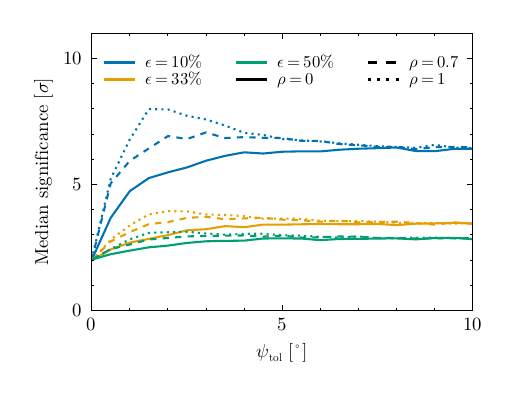}
    \caption{\textbf{\textit{Significance as a function of $\psi_{\mathrm{tol}}$ for alternative signal case.}}
    This shows the same data as \cref{fig:significance} but with $\gamma_{\mathrm{sig}} = 2.0$, $\gamma_{\mathrm{bg}} = 3.7$, and signal counts tuned to give $\sim 2\,\sigma$ at $\psi_{\mathrm{tol}} = 0$.
    The qualitative improvement from the quality-dependent selection is robust to the choice of signal spectral index.
    }
    \label{fig:spectral_scan}
\end{figure}

To verify that the sensitivity improvement is not specific to a particular source spectrum, we repeat the scan with a harder signal spectral index $\gamma_{\mathrm{sig}} = 2.0$---compared to $\gamma_{\mathrm{sig}} = 3.2$ previously.
The signal counts are adjusted so that the baseline significance at $\psi_{\mathrm{tol}} = 0$ is $\sim 2\,\sigma$.

\cref{fig:spectral_scan} shows the result.  The qualitative behavior is unchanged: significance increases with tolerance up to a broad optimum, and the improvement is larger for smaller $\epsilon$ and larger $\rho$.
The harder spectrum concentrates signal events at higher energies where the angular resolution is narrower, yet the quality-dependent selection remains effective.
This validates that our procedure is robust to different spectral indices and gives improvement even when a source does not already have a high significance.
This latter point implies that this technique can be applied to catalog searches, the computational implications of which we discuss further in \cref{app:computation}.

\textbf{\textit{Discussion and Outlook---}}We have proposed a simple yet effective strategy for improving neutrino point source searches: preferentially retaining events whose early-level reconstruction falls near candidate source directions for higher-quality reconstruction, while continuing to subsample the remaining events at the baseline computational budget.
This source-informed event selection exploits the growing catalog of neutrino source candidates to enhance sensitivity where it matters most.

Our results demonstrate that this approach can improve median significance by factors of $\sim 2$--$3$ compared to uniform subsampling, with the improvement depending on baseline efficiency, angular tolerance, and inter-level quality correlation.
Even in the most conservative case we consider, the significance grows by 25\%, which corresponds to more than a 50\% increase in exposure for background-dominated analyses.
Crucially, these gains come with modest computational overhead: for catalogs of $\mathcal{O}(100)$ sources, the additional $L_2$ processing amounts to $\sim 7$--$14\%$ at typical baseline efficiencies.
These improvements are significant, given the projected improvements of using traditional methods with all next-generation neutrino detectors~\cite{Schumacher:2025qca}.

The method is robust across different source spectra and remains effective even when baseline significance is low ($\sim 2\sigma$), making it applicable to catalog searches and marginal source candidates.
Furthermore, the source scrambling technique we validate in \cref{fig:pdfs} allows experimenters to preserve the traditional right-ascension randomization approach for background estimation while implementing directional selection.

% The gains demonstrated here are largest in the low-efficiency regime (\cref{fig:significance}), suggesting that the most immediate experimental targets are high-purity, low-statistics event selections rather than large-throughput muon-track samples.
% Samples such as HESE~\cite{IceCube:2020wum}, ESTES~\cite{IceCube:2024yzw}, and DNNCascades~\cite{IceCube:2023ame} operate with stringent quality criteria that impose low effective efficiencies, placing them squarely in the regime where our method yields the greatest improvement.
% ESTES provides a particularly concrete illustration: events are currently rejected if their early-level reconstruction is geometrically consistent with an atmospheric muon traversing a gap in the instrumented volume without leaving hits.
% A source-informed variant could relax this veto for corridors aligned with a known source candidate, retaining events that would otherwise be discarded.
% A similar argument applies to the PSTracks sample~\cite{IceCube:2021xar} in the southern sky, where an elevated energy threshold is imposed to suppress the atmospheric muon background; a source-informed selection could lower the effective threshold for events reconstructed near candidate source directions, extending sensitivity to lower signal energies.

Several extensions merit consideration.
First, while we have focused on two-level reconstruction chains, many experiments employ three or more levels.
The method generalizes naturally: at each level, events consistent with source directions are preferentially retained for the next level.
One may also consider different criteria for determining the probability that an event comes from a known source.
For instance, when using likelihood-based approaches, one could evaluate the likelihood difference between the best-fit point and the source position and retain any events with a sufficiently small difference.

Looking ahead, the next generation of neutrino telescopes---IceCube-Gen2, P-ONE, KM3NeT---will collect unprecedented event rates, making computational budgets for high-level reconstruction increasingly constrained.
Simultaneously, multi-messenger astronomy is identifying an increasing number of candidate neutrino sources from electromagnetic and gravitational-wave observations.
Additionally, next-generation observatories, such as TAMBO~\cite{TAMBO:2025jio}, TRINITY~\cite{Otte:2025dld}, and HERON~\cite{GRAND:2025rps}, will provide high-purity catalogs of event locations that can seed this process.
Source-informed event selection provides a path to exploit both developments: it uses the known source catalog to intelligently allocate limited computational resources, enhancing discovery potential without requiring new detector capabilities or reconstruction algorithms.

As neutrino astronomy transitions from discovery-driven to catalog-driven science, data processing strategies must evolve accordingly.
The source-agnostic event selection that served us well in the early era of neutrino telescopes is no longer optimal.
By incorporating directional information from known sources into our event selection, we can substantially improve our sensitivity to the sources we care most about discovering.

\textbf{\textit{Acknowledgements---}}
We are thankful for the engaging and constructive discussions with Naoko Kurahashi Neilson, Chad Finley, Ali Kheirandish, and Will Luszczak.
CAA and JL thanks the organizers of the Neutrino2024 conference for providing a venue suitable for discussions and the commune of Ixelles for maintaining their public spaces, where some of the conversations of this work took place.
Additionally, CAA thanks the organizers of the \textit{Extragalactic and Galactic Neutrino Astronomy} workshop, where engaging discussions associated with this work took place.
This work was made possible by the Hoover Seedfund between Harvard University and University Catholique de Louvain, which facilitated discussions between CA and JL.
JL is supported by the Belgian Fonds de la Recherche Scientifique (FRS-FNRS).
CAA are supported by the Faculty of Arts and Sciences of Harvard University, Canadian Institute for Advanced Research (CIFAR), the National Science Foundation (NSF), the Research Corporation for Science Advancement, and the David \& Lucile Packard Foundation.
PZ is funded by the David \& Lucile Packard Foundation through this work.

\bibliography{references}
\clearpage

\begin{appendix}
\section{Supplemental Material}

\subsection*{Quality Dependent Cuts}
\refstepcounter{section}
\label{app:other_cases}

In the main analysis, the inter-level selection retains all events whose $L_1$ reconstruction falls within the cone and randomly subsamples the remainder at rate $\epsilon$.
In reality, higher-quality events are more likely to be retained through the event selection.
One might worry that this would erode the gains from the cone-based selection.

\begin{figure}[t]
    \centering
    \includegraphics[width=\columnwidth]{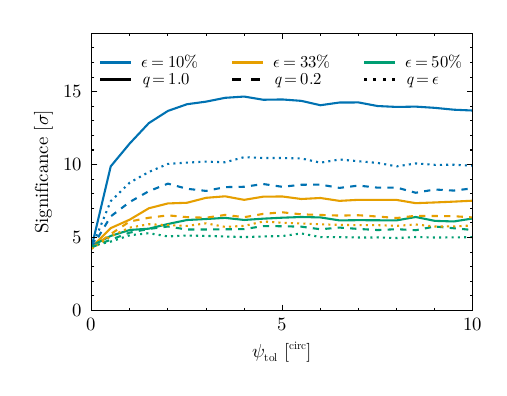}
    \caption{\textbf{\textit{Significance versus $\psi_{\mathrm{tol}}$ for quality choices.}}
    Asimov significance versus angular tolerance for $\rho = 0.7$, comparing no quality bias (solid), $q = 0.2$ quality bias (dashed), and $q = \epsilon$ quality bias (dotted).  Colors indicate baseline efficiency: $\epsilon = 10\%$ (blue), $33\%$ (green), $50\%$ (red).
    Quality-biased selection reduces the absolute gain from the cone but does not eliminate it.
    }
    \label{fig:quality_bias}
\end{figure}

To study this, we introduce a quality-biased selection that unconditionally retains events in the top $q$ fraction of $L_1$ reconstruction quality (as measured by the $L_1$ angular error quantile), in addition to the cone and random selection.
An event is retained if \emph{any} of the three conditions is satisfied: (i) $L_1$ falls within the cone, (ii) $L_1$ quality is in the top $q$ fraction, or (iii) the event is randomly kept at rate $\epsilon$.

\cref{fig:quality_bias} shows the significance as a function of tolerance for $\rho = 0.7$, comparing three scenarios: no quality bias ($q = 0$, solid), moderate quality bias ($q = 0.2$, dashed), and efficiency-matched quality bias ($q = \epsilon$, dotted).
Colors distinguish the baseline efficiency $\epsilon$.

The cone-based selection continues to improve sensitivity in all cases,
though the absolute gain is reduced when quality bias is present.  This is
expected: the quality-biased events consume part of the computational
budget that would otherwise be available for cone-selected events,
diluting the directional enrichment.  The effect is most pronounced at
low $\epsilon$, where each additional non-cone event has a larger
relative impact.  Nevertheless, even with aggressive quality bias
($q = \epsilon$), the cone selection provides a meaningful improvement
over the $\psi_{\mathrm{tol}} = 0$ baseline.

\subsection*{Fit validation}
\refstepcounter{section}
\label{app:fit_validation}

To verify that the likelihood fitter correctly recovers the injected signal
strength, we perform a closure test: for each of two representative
configurations, we generate Poisson realizations with a known
$N_{\mathrm{sig}}$ and compare the fitted $\hat{n}_s$ to the true value
across many trials.

\begin{figure}[t]
    \centering
    \includegraphics[width=\columnwidth]{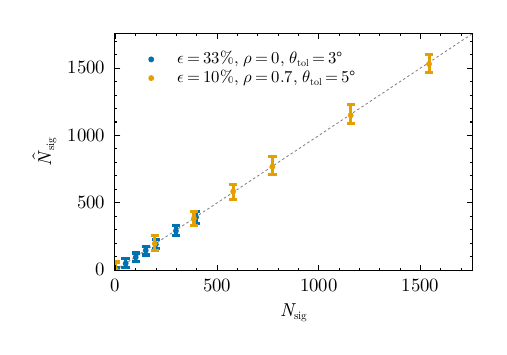}
    \caption{\textbf{\textit{Fitted signal count $\hat{n}_s$ versus injected signal.}}
    count $N_{\mathrm{sig}}$ for two representative configurations.
    Points show the median over 200 trials; error bars indicate the central 68\% interval.
    The dashed line shows $\hat{n}_s = N_{\mathrm{sig}}$.
    }
    \label{fig:fit_validation}
\end{figure}

Under the background-only hypothesis, the test statistic given by \cref{eq:ts} is expected to follow a mixture distribution:
$\frac{1}{2}\delta(0) + \frac{1}{2}\chi^2(1)$, where the point mass at
zero corresponds to realizations in which the best-fit signal count is at
the physical boundary $\hat{n}_s = 0$~\cite{Cowan:2010js}.

\begin{figure}[t]
    \centering
    \includegraphics[width=\columnwidth]{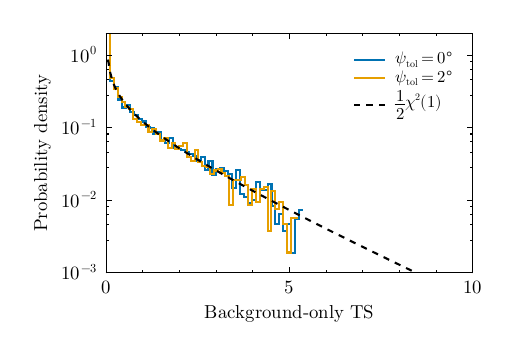}
    \caption{\textbf{\textit{Background-only TS distribution.}}
    This is computed 10,000 trials with $\epsilon = 33\%$,
    $\rho = 0$ at $\psi_{\mathrm{tol}} = 0^\circ$ and
    $\psi_{\mathrm{tol}} = 2^{\circ}$.
    The dashed line shows the theoretical $\frac{1}{2}\chi^2(1)$ prediction for the positive TS
    values.}
    \label{fig:bg_ts}
\end{figure}

\cref{fig:bg_ts} shows the TS distribution from 10{,}000 background-only
pseudo-experiments for $\epsilon = 33\%$, $\rho = 0$, at two tolerance
values.  The positive TS values are well described by the
$\frac{1}{2}\chi^2(1)$ envelope, confirming the validity of the test
statistic calibration.

\subsection{Computational Overhead}
\refstepcounter{section}
\label{app:computation}

A natural concern is the additional computational cost of the quality-dependent selection when applied to multiple source candidates simultaneously.
The fraction of events requiring $L_2$ reconstruction
under the selection protocol is
\begin{equation}
    f_{\mathrm{L2}} = N_{\mathrm{src}}\, f_{\mathrm{cone}}
                    + \bigl(1 - N_{\mathrm{src}}\, f_{\mathrm{cone}}\bigr)\,\epsilon,
\end{equation}
where $f_{\mathrm{cone}} = (1 - \cos\psi_{\mathrm{tol}})/2$ is the
fractional solid angle of each cone and $N_{\mathrm{src}}$ is the number
of source candidates (assuming non-overlapping cones).  Without the
selection, the fraction is simply $\epsilon$.  The fractional increase in
$L_2$ processing relative to uniform subsampling is
$N_{\mathrm{src}}\, f_{\mathrm{cone}}\,(1 - \epsilon) / \epsilon$,
which is dominated by the isotropic background events whose $L_1$ reconstruction happens to fall within a cone and would not otherwise have been selected.
\cref{tab:overhead} evaluates this for
$\psi_{\mathrm{tol}} = 3^\circ$.
For a catalog of $\mathcal{O}(100)$ sources, the overhead is modest: $\sim 14\%$ at $\epsilon = 33\%$ and $\sim 7\%$ at $\epsilon = 50\%$.

\begin{table}[h]
    \centering
    \caption{Fractional increase in $L_2$ processing load relative to
    uniform subsampling at rate $\epsilon$, for
    $\psi_{\mathrm{tol}} = 3^\circ$
    ($f_{\mathrm{cone}} = 6.85 \times 10^{-4}$).}
    \label{tab:overhead}
    \begin{ruledtabular}
    \begin{tabular}{r c c c}
        $N_{\mathrm{src}}$ & $\epsilon = 10\%$ & $\epsilon = 33\%$
                           & $\epsilon = 50\%$ \\
        \hline
        1    & $+0.6\%$   & $+0.1\%$   & $+0.07\%$  \\
        10   & $+6.2\%$   & $+1.4\%$   & $+0.7\%$   \\
        50   & $+30.8\%$  & $+6.9\%$   & $+3.4\%$   \\
        100  & $+61.7\%$  & $+13.7\%$  & $+6.9\%$   \\
        500  & $+308\%$   & $+68.5\%$  & $+34.3\%$  \\
    \end{tabular}
    \end{ruledtabular}
\end{table}

\end{appendix}
\end{document}